\overfullrule=0pt
\input harvmac

\def\plb{{+\bar l}}
\def\mlb{{-\bar l}}

\def\pl{{+ l}}
\def\p{{\partial}}
\def\pb{{\bar\partial}}
\def\ml{{- l}}
\def\mm{{- m}}
\def\half {{1 \over 2}}
\def\k{{\kappa}}
\def\S{{\Sigma}}
\def\d{{\delta}}
\def\T{{\Theta}}
\def\Tb{{ \bar\Theta}}
\def\t{{\theta}}
\def\e{{\epsilon}}
\def\l{{\lambda}}

\lref\amati{
D.~Amati and C.~Klimcik,
{\it
Nonperturbative Computation 
Of The Weyl Anomaly For A Class Of Nontrivial Backgrounds,}
Phys.\ Lett.\ B {\bf 219}, 443 (1989).
}

\lref\horsteif{
G.~T.~Horowitz and A.~R.~Steif,
{\it Space-Time Singularities In String Theory,}
Phys.\ Rev.\ Lett.\  {\bf 64}, 260 (1990).
}

\Title{\vbox{\hbox{IFT-P.051/2002 }}}
{\vbox{
\centerline{\bf N=2 Superconformal Description of Superstring}
\centerline{\bf in Ramond-Ramond Plane Wave Backgrounds}}}
\bigskip
\centerline{Nathan Berkovits\foot{e-mail: nberkovi@ift.unesp.br}}
\smallskip
\centerline{\it Instituto de F\'\i sica Te\'orica, Universidade Estadual
Paulista}
\centerline{\it Rua Pamplona 145, 01405-900, S\~ao Paulo, SP, Brasil}

\bigskip
\centerline{Juan Maldacena\foot{e-mail: malda@ias.edu}}
\smallskip
\centerline{\it Institute for Advanced Study}
\centerline{\it Princeton, NJ 08540, USA}

\vskip .3in

Using the U(4) formalism developed ten years ago, the 
worldsheet action for the superstring in Ramond-Ramond plane wave
backgrounds is expressed in a manifestly N=(2,2) superconformally invariant
manner. This simplifies the construction of consistent Ramond-Ramond
plane wave backgrounds
and eliminates the problems associated with light-cone interaction point
operators.

\Date {August 2002}

\newsec{Introduction}

Although the construction of worldsheet actions for the superstring
in Neveu-Schwarz backgrounds is well understood using the Ramond-Neveu-Schwarz
(RNS) formalism, the worldsheet
action for the superstring in Ramond-Ramond (RR) backgrounds has been less
studied. Because of the important role of Ramond-Ramond backgrounds
in the AdS/CFT conjectures \ref\mone{J.M. Maldacena,
{\it The Large N Limit of Superconformal Field Theories and Supergravity},
Adv. Theor. Math. Phys. 2 (1998) 231, Int. J. Theor. Phys. 38 (1999) 1113,
hep-th/9711200.}, these worldsheet actions might be very
useful for studying aspects of these conjectures.

When the background allows a light-cone gauge choice,
the most straightforward method for constructing the superstring action
in Ramond-Ramond backgrounds is to use the light-cone Green-Schwarz
(GS) formalism \ref\mets{R.R. Metsaev, {\it Type II Green-Schwarz 
Superstring in Plane Wave Ramond-Ramond Background}, Nucl. Phys. B625
(2002) 70, hep-th/0112044.}. This light-cone formalism is extremely useful for
computing the physical spectrum in a given background%
, however, it
is difficult to use for computing scattering amplitudes or for 
determining consistency conditions on superstring backgrounds.
Although the light-cone RNS formalism suffers from similar problems,
in the RNS case there exists an N=1 superconformally invariant description
of the superstring which can replace the light-cone description (at
least when RR fields are zero).

Over ten years ago, an N=2 superconformally invariant description of
the superstring was proposed \ref\het{N. Berkovits, {\it
The Heterotic Green-Schwarz Superstring on an N=(2,0) Worldsheet},
Nucl. Phys. B379 (1992) 96, hep-th/9201004.}\ref\twist{
N. Berkovits, {\it Calculation of Green-Schwarz Superstring Amplitudes
using the N=2 Twistor-String Formalism}, Nucl. Phys. B395 (1993) 77,
hep-th/9208035.}
which reduces in light-cone gauge
to the light-cone GS description. In addition to the light-cone
transverse variables which are combined into bosonic
worldsheet N=2 superfields, this superconformally invariant description
also contains fermionic worldsheet N=2 superfields which describe
the longtitudinal variables. In a flat background, the action is
quadratic and contains manifest U(4) Lorentz invariance as well as
twenty manifest spacetime supersymmetries. This U(4) formalism can
be obtained by gauge-fixing a ``doubly-supersymmetric'' action
\ref\double{
D. Sorokin, V. Tkach, D. Volkov and A. Zheltukhin,
{\it From the Superparticle Siegel Symmetry to the Spinning Particle
Proper Time Supersymmetry}, Phys. Lett. B216 (1989) 302\semi
N. Berkovits, {\it A Covariant Action for the Heterotic Superstring
with Manifest Spacetime Supersymmetry and Worldsheet Superconformal
Invariance}, Phys. Lett. B232 (1989) 184\semi
D. Sorokin, {\it Superbranes and Superembeddings}, Phys. Rept. 329 (2000)
1, hep-th/9906142.}\ref\tontwo{
M. Tonin, {\it Worldsheet
Supersymmetric Formulations of Green-Schwarz Superstrings}, Phys. Lett.
B266 (1991) 312.}\het\ where $\kappa$-symmetry is replaced by
worldsheet supersymmetry, and is related to the
RNS formalism by a field redefinition which maps the fermionic
N=2 superconformal generators to the twisted BRST current and $b$
ghost in the RNS formalism \ref\twistrns{N. Berkovits,
{\it The Ten-Dimensional Green-Schwarz Superstring is a Twisted 
Neveu-Schwarz-Ramond String}, Nucl. Phys. B420 (1994) 332\semi
N. Berkovits and C. Vafa, {\it On the Uniqueness of String Theory},
Mod. Phys. Lett. A9 (1994) 653, hep-th/9310170\semi
N. Berkovits and C. Vafa, {\it N=4 Topological Strings}, Nucl. Phys.
B433 (1995) 123, hep-th/9407190.}. 
Recently, this U(4) formalism has
also been related to the pure spinor formalism \ref\puresp
{N. Berkovits, {\it Super-Poincar\'e Covariant Quantization of the
Superstring}, JHEP 0004 (2000) 018, hep-th/0001035.} for the superstring
via different gauge fixings 
\ref\tonin
{M. Matone, L. Mazzucato, I. Oda, D. Sorokin and M. Tonin, {\it
The Superembedding Origin of the Berkovits Pure Spinor Covariant
Quantization of Superstrings}, hep-th/0206104.}
of the doubly-supersymmetric action. 

In the pure spinor formalism of the superstring \puresp,
all $SO(9,1)$ super-Poincar\'e invariance is manifest 
but the original N=2
worldsheet superconformal invariance is hidden. Although one can
use the pure spinor formalism to describe any consistent background
of the superstring, for
describing Ramond-Ramond plane wave backgrounds in which Lorentz
invariance is already broken, it is more advantageous
to use the U(4) formalism so that the N=2
worldsheet supersymmetry is manifest. For example, the conformally
invariant action for the maximally
supersymmetric Ramond-Ramond plane wave background 
\ref\blau {M. Blau,
J. Figueroa-O'Farrill, C. Hull and G. Papadopoulos,
{\it A New Maximally Supersymmetric Background of
IIB Superstring Theory}, JHEP 0201 (2002) 047,
hep-th/0110242.} will be much simpler
using the U(4) formalism than using the pure spinor formalism
\ref\confps{N. Berkovits, {\it Conformal Field Theory for the
Superstring in a Ramond-Ramond Plane Wave Background}, JHEP 0204
(2002) 037, hep-th/0203248.}.
As will be shown here,
the U(4) formalism
is also
extremely useful for determining consistency conditions for
Ramond-Ramond plane wave backgrounds with
fewer numbers of supersymmetries
\ref\mtwo{
J.~Maldacena and L.~Maoz,
{\it Strings on pp-Waves and Massive Two Dimensional Field Theories,}
arXiv:hep-th/0207284. }.
Furthermore,
since this formalism does not require light-cone interaction point
operators, it does not suffer from contact term problems and may
also be useful 
for computing scattering amplitudes in these backgrounds.\foot{In
\twist\ and
\ref\ampli{N. Berkovits, {\it Finiteness and Unitarity of Lorentz-Covariant
Green-Schwarz Superstring Amplitudes}, Nucl. Phys. B408 (1993) 43,
hep-th/9303122.}, multiloop scattering amplitudes were computed using
the U(4) formalism and a proof was claimed for their
finiteness. This proof was incorrect because of subtleties coming from
unphysical poles in multiloop correlation functions of chiral bosons.
However, the techniques developed in these papers can be
used for computing tree and one-loop scattering
amplitudes where such subtleties are absent.}

In Section 2 of this paper, the U(4) formalism and its
relation to the light-cone GS formalism will be reviewed. In
Section 3, N=(2,2) superconformally invariant actions will
be constructed for the superstring in plane wave
Ramond-Ramond backgrounds which
preserve either two or four spacetime supersymmetries, and will be proven
to be exact superconformal field theories to all perturbative orders in
$\alpha'$.
Such Ramond-Ramond
backgrounds are naturally described by real or holomorphic 
superpotentials \mtwo and may be useful for studying various aspects of
AdS/CFT conjectures. And in the Appendix, the relation between
N=2 superconformal invariance and the light-cone GS interaction point
operator will be discussed.

\newsec{Review of U(4) Formalism}

\subsec{Flat background}

In a flat background, the Type IIB GS action in
light-cone gauge can
be written as \ref\gslc{M.B. Green and
J.H. Schwarz, {\it Supersymmetrical Dual String Theory},
Nucl. Phys. B181 (1981) 502.}
\eqn\actone{S=\int d^2 z  ( \p x^\plb \pb x^\ml +s^\mlb
\pb s^\pl + \bar s^\mlb \p \bar s^\pl ) }
where $SO(8)$ has been broken to $SU(4)\times U(1)$
such that the $SO(8)$ vector splits into 
$(x^\plb,x^\ml)$ and the $SO(8)$ chiral spinor splits
into $(s^\mlb, s^\pl)$ for $l=1$ to 4. Note that under
$SU(4)\times U(1)$, $x^\plb$ and $x^\ml$ transform as
$(\bar 4,+1)$ and $(4,-1)$ representations, whereas
$s^\mlb$ and $s^\pl$ transform as
$(\bar 4,-1)$ and $(4,+1)$ representations. The action of
\actone\ can easily be written in N=2 supersymmetric
notation as 
\eqn\acttwo{S=\int d^2 z \int d^2 \k^+ \int d^2 \k^- X^\plb
X^\ml }
where $X^\plb$ and $X^\ml$ are chiral and antichiral superfields
satisfying 
\eqn\chiralX{D_- X^\plb = \bar D_- X^\plb=0, \quad
D_+ X^\ml = \bar D_+ X^\ml=0,}
$$D_-= {\p\over{\p\k^-}} +\k^+ \p_z,\quad
D_+= {\p\over{\p\k^+}} +\k^- \p_z,\quad
\bar D_-= {\p\over{\p\bar\k^-}} +\bar\k^+ \pb_{\bar z},\quad
\bar D_+= {\p\over{\p\bar\k^+}} +\bar\k^- \pb_{\bar z},$$
\eqn\compX{X^\plb = x^\plb + \k^+ s^\mlb +\bar\k^+ \bar s^\mlb
+ \k^+\bar \k^+ h^\mlb,\quad
X^\ml = x^\ml + \k^- s^\pl +\bar\k^- \bar s^\pl
+ \k^-\bar \k^- h^\pl,}
and $(h^\mlb, h^\pl)$ are auxiliary fields.

To construct a conformally invariant action which reduces
in light-cone gauge to \actone, it is useful to recall the construction
of conformally invariant actions for the bosonic and RNS string.
In bosonic string theory, we can find a map from the complex plane
to 
the light-cone string diagram
$\rho(z)$ which is a conformal transformation.
To construct a conformally invariant action for the bosonic string,
the real part of this conformal map $\rho(z) + \bar \rho(\bar z)$
 is promoted to 
the target-space light-cone variable $x^+(z,\bar z)$.
A similar procedure can be performed for RNS string theory, however,
in this case the light-cone string diagram is mapped to an N=1
complex superplane using the N=1 superconformal map 
$[\rho(z,\k), \xi(z,\k)]$ where $[\rho,\xi]$ parameterize the
string diagram and $[z,\k]$ parameterize the N=1 superplane.
As reviewed in the Appendix,
the use of an N=1 superconformal map is crucial for obtaining
the appropriate light-cone RNS interaction point operator.
To construct an N=1 superconformally invariant action,
one promotes $\rho(z,\kappa) + \bar \rho(\bar z, \bar \kappa)$ to the 
worldsheet N=1 superfield $X^+$. Since the map is $N=1$ superconformal,
$\xi(z,\k)$ is determined by
$\rho(z,\k)$, so $X^+$ completely determines the N=1
superconformal map.

It turns out 
that to obtain a conformally
invariant description of the GS superstring, one needs
to use an N=2 superconformal
transformation 
$$[\rho(z,\k^+,\k^-),\xi^+(z,\k^+),\xi^-(z,\k^-)]$$ 
to map the N=2 complex superplane
$[z,\k^+,\k^-]$ to the string diagram $[\rho,\xi^+,\xi^-]$. 
In fact, as was shown in \het\ and reviewed in the Appendix, 
precisely
such an N=2 superconformal map is required for obtaining the
correct light-cone GS interaction point operator.
However, unlike N=1 superconformal transformations, 
the $\xi^+(z,\k^+)$ and 
$\xi^-(z,\k^-)$ parameters in an N=2 superconformal map are
not uniquely determined by $\rho(z,\k^+,\k^-)$ because of the
possibility of performing $U(1)$ transformations.
For this reason, instead of promoting one variable to a
 superfield we 
actually need to promote two. 
We promote 
$\xi^+(z,\k^+)$ and
$\xi^-(z,\k^-)$ to worldsheet superfields
which will be called $\T^+$ and 
$\T^-$.  For the Type II superstring, one also has the
barred fermions $\bar\xi^+(\bar z,\bar\k^+)$ and
$\bar\xi^-(\bar z,\bar\k^-)$ which will be promoted to
$\bar\T^+$ and $\bar\T^-$. 

These fermionic target-space variables will be defined
as N=2 chiral and antichiral worldsheet
superfields satisfying \het\twist
\eqn\chiralT {D_-\T^+ = \bar D_-\T^+=   
D_-\Tb^+ = \bar D_-\Tb^+=0,\quad
D_+\T^- = \bar D_+\T^-=
D_+\Tb^- = \bar D_+\Tb^-=0.}
 The fact that the map $[z,\kappa^-,\kappa^-] \to [\rho, \xi^+ ,\xi^-]$
 is $N=2$ superconformal then 
determines $\rho(z,\kappa^+,\kappa^-)$ which in turn determines
$X^+$. More explicitly, we have 
that 
 $\rho$ is related to $\xi^\pm$ by
$\xi^+ D_- \xi^- =D_-\rho$ and
$\xi^- D_+ \xi^+ =D_+\rho$. The action will be chosen in 
such a way that the equations of motion for $\T^\pm ,~\bar \T^\pm$ are
\eqn\eqmotion{
\bar D_- \T^- = \bar D_+ \T^+ = D_- \bar \T^- = D_+ \bar \T^+ =0
.}
This implies, in particular, that $\T^\pm$  are holomorphic.
For any solution of the equations of motion we define $X^+$ through
\eqn\restr{
\T^+ D_- \T^- =D_- X^+, \quad
\T^- D_+ \T^+ =D_+ X^+, }
$$
\bar\T^+ \bar D_- \bar\T^- =\bar D_- X^+, \quad
\bar \T^- \bar D_+ \bar \T^+ =\bar D_+ X^+.$$
Since \restr\ determines $X^+$ up to a constant shift in terms
of $\T^\pm$ and $\bar\T^\pm$, one can treat $\T^\pm$ and $\bar\T^\pm$
as the fundamental target-space variables and treat $X^+$ as a composite
variable. The only subtlety is that on worldsheets with non-zero genus,
the equations
\eqn\subtle{[D_+,D_-] (\T^+\T^-) = \p X^+,\quad
[\bar D_+,\bar D_-] (\bar \T^+\bar \T^-) = \bar \p X^+}
imply that $(\T^\pm,\bar\T^\pm)$ must satisfy 
\eqn\loops{\int dz \int d\k^+\int d\k^- \T^+\T^- = 
\int d\bar z \int d\bar \k^+\int d\bar \k^- \bar \T^+\bar \T^- }
when integrated around a non-contractible loop. In other words, these
constraints come from the fact that $X^+$ defined through \restr\ should
be single valued. 

To construct a worldsheet action, one needs conjugate momenta for
$(\T^\pm,\bar\T^\pm)$ which will be defined as the partially chiral
superfields $(W^\pm, \bar W^\pm)$ restricted to satisfy 
\eqn\chiralW{\bar D_- W^+ = 0,\quad\bar D_+ W^-=0,
\quad D_-\bar W^+=0,\quad D_+ \bar W^-=0.} With the constraints
of \chiralX, \chiralT\ and \chiralW, one can construct the critical N=(2,2)
superconformally invariant action
\eqn\actthree{S=\int d^2 z \int d^2 \k^+ \int d^2 \k^- (X^\plb
X^\ml  + W^+\T^- + W^-\T^+ + \bar W^+\bar \T^- + \bar W^-\bar \T^+),}
which is such that the equations of motion for $W^\pm, ~\bar W^\pm$
enforce \eqmotion . In components \actthree\ becomes
\eqn\actfour{\eqalign{
S= & \int d^2 z  ( \p x^\plb \bar\p x^\ml + s^\mlb\bar\p s^\pl
+\bar s^\mlb \p\bar s^\pl + h^\mlb h^\pl  +
 p^+\bar\p\t^- +p^-\bar\p\t^+ + \cr & + \bar p^+\p\bar\t^- 
+\bar p^-\p\bar\t^+
+  w^+\bar\p\l^- +w^-\bar\p\l^+ + \bar w^+\p\bar\l^- 
+\bar w^-\p\bar\l^+ ),
}}
where $X^\plb$ and $X^\ml$ are defined in \compX, 
\eqn\compex{\T^\pm =\t^\pm+ \k^\pm \l^\pm + ... , \quad
\bar\T^\pm =\bar\t^\pm+ \bar\k^\pm \bar\l^\pm + ... , }
$$W^\pm = \k^\pm w^\pm + p^\pm \k^+\k^-  + ... , \quad
\bar W^\pm = \bar \k^\pm \bar w^\pm + 
\bar p^\pm \bar \k^+\bar \k^-  + ... , $$
and $\dots $ includes auxiliary fields which have been ignored in the
action of \actfour.
The left-moving N=2 stress tensor for this action is given by the superfield
\eqn\twostress{T= D_+ W^+ D_-\T^- - D_- W^- D_+\T^+ + D_+ X^\plb
D_- X^\ml,}
which contains critical N=2 central charge since $[X^\plb,X^\ml]$
contribute $c=12$ and
$[\T^\pm,W^\mp]$ contribute $c=-6$.

Like the light-cone action of \acttwo, the action of \actthree\
is manifestly invariant under $SU(4)\times U(1)$ Lorentz
transformations which transform $[X^\plb,X^\ml, \T^\pm,\Tb^\pm,
W^\pm,\bar W^\pm]$ as 
$[\bar 4_1,4_{-1}, 1_{\pm 2}, 1_{\pm 2}, 1_{\pm 2}, 1_{\pm 2}]$ 
representations where the subscript denotes the $U(1)$ charge.
However, in addition to the sixteen manifest light-cone spacetime
supersymmetries, the action of \actthree\ is
also manifestly invariant under four additional spacetime
supersymmetries. Under these twenty spacetime supersymmetries 
parameterized by $[\e^{+l},\e^{-\bar l}, \bar\e^{+l},\bar\e^{-\bar l},
\e^\pm, \bar\e^\pm]$, the worldsheet variables of \actthree\ transform
as 
\eqn\susytr{\d X^\plb=\e^{-\bar l} \T^+ +\bar\e^{-\bar l}\bar\T^+,\quad
\d X^\ml=\e^{+ l} \T^- +\bar\e^{+ l}\bar\T^-,}
$$\d\T^\pm = \e^\pm, \quad
\d\bar\T^\pm = \bar\e^\pm, $$
$$\d W^+ =-\e^{+ l} X^\plb,\quad
\d W^- =-\e^{-\bar l} X^\ml,\quad
\d \bar W^+ =-\bar\e^{+ l} X^\plb,\quad
\d \bar W^- =-\bar\e^{-\bar l} X^\ml.$$
One can check that the supersymmetry transformations of \susytr\
anticommute to give translations. The only subtlety is that
translations in the $x^+$ direction leave the worldsheet variables
$\T^\pm$ and $W^\pm$ invariant since \restr\ implies that they are
independent of the $x^+$ zero mode. And translations in the $x^-$
direction transform $\d W^\pm = c\T^\pm$ and $\d\bar W^\pm= c\Tb^\pm$,
as can be seen from the translation generator
$P^+ = \int dz \int d^2\k \T^+\T^- + \int d\bar z \int d^2 \bar \k
\bar\T^+\bar \T^-$.
 
\subsec{Consistency of light-cone background}

It is clear by construction that the action \actthree\  reduces to 
\acttwo\ in lightcone gauge. Nevertheless, let us see this more
explicitly. The equations of motion for $\T^\pm$  \eqmotion\
 together with the chirality constraints \chiralT\ imply
that $\T^\pm$ are holomorphic functions which  via an $N=2$ 
superconformal transformation can be set to $\T^\pm = \kappa^\pm $.
\restr\ then implies  that $\partial x^+ = 1$. 

In a non-flat background, the action of 
\actthree\ is replaced by a non-quadratic action which can
depend in a complicated manner on the worldsheet superfields.
Since the fermionic N=2 superconformal generators are related by
a field redefinition to the BRST current and $b$ ghost in the
RNS formalism, one expects that quantum N=(2,2) superconformal
invariance of the action implies that background is an exact
solution of superstring theory. As will now be argued, this condition
of N=2 superconformal invariance can be used to determine when a
given light-cone background in the GS formalism describes a solution of
superstring theory.

Suppose one is given an action depending on the light-cone GS variables
$X^\plb$ and $X^\ml$ of \chiralX.  
In general, the action will not be N=2 superconformally invariant or
even N=2 worldsheet supersymmetric. However, by coupling $\T^\pm$ and
$\bar\T^\pm$ in an appropriate manner, one can always construct an
action which is classically N=(2,2) superconformally invariant and
which reduces to the original action in light-cone gauge where
$\l^\pm=\bar\l^\pm=1$ and $\t^\pm=\bar\t^\pm=0$. If this new action
is also N=(2,2) superconformally invariant at the quantum level, then
the original light-cone GS background describes a solution of
superstring theory. Note that an analogous construction exists for
bosonic and RNS light-cone backgrounds where, for the bosonic string,
$\p x^+$ and $\bar\p x^+$ are used to construct conformally invariant
actions and, for the RNS superstring, $DX^+$ and $\bar DX^+$ are
used to construct N=(1,1) superconformally invariant actions.

Because the parametrization of $X^+$ through 
\restr\ closely resembles the twistor
constraints described in \double, the action of \actthree\ has been
called the N=2 twistor-string action.
Although this 
action is not manifestly Lorentz invariant,
it can be related to the manifestly Lorentz-invariant ``doubly-supersymmetric'' 
action of \tontwo\het\
by introducing additional gauge and auxiliary fields.
Furthermore, this doubly-supersymmetric action 
has been recently related in \tonin\
to the ``pure spinor'' formalism for the superstring \puresp\ 
in which the superstring is 
quantized in a manifestly
super-Poincar\'e invariant manner by constructing a BRST operator out
of pure spinors. Also, the U(4)-invariant action of \actthree\  
can be
related to the standard Lorentz-invariant RNS worldsheet action of 
\ref\fms{D. Friedan, E. Martinec and S. Shenker, 
{\it Conformal Invariance, Supersymmetry and String Theory},
Nucl. Phys. B271 (1986) 93.} 
by bosonizing some
of the worldsheet fields and interpreting the resulting theory as
an $N=1 \to N=2$ ``embedding'' of the RNS superstring \twistrns. However,
none of these Lorentz-invariant descriptions of the superstring
preserve manifest
N=2 worldsheet superconformal invariance. As will be shown in the
following section, 
for Ramond-Ramond plane wave backgrounds in which
Lorentz invariance is already broken, the most convenient
description is the U(4) formalism 
which preserves
manifest N=2 superconformal
invariance.

\newsec{U(4) Formalism for Plane Wave Background}

In an plane wave background, the target-space fields are independent
of $x^-$ so that the equations of motion for $x^+$ are $\p\bar\p x^+=0$.
Since $x^-$ is contained in the superfields $W^\pm$ and $\bar W^\pm$
in the U(4) formalism, the plane wave background fields should
be independent of $W^\pm$ and $\bar W^\pm$. So the most general 
classically
N=2 superconformal invariant action for a plane wave background is 
\eqn\actfive{S= S_0 + \int d^2 z d^4\k  ~
U(X^\plb,X^\ml,\T^-,\T^+,\Tb^+,\Tb^-)}
where $S_0$ is the action in a flat background of \actthree\ and 
$U$ is a general scalar superfield.
The left-moving N=2 stress tensor in this background is 
\eqn\backstress{T=T_0 + D_+ X^\plb D_- X^\mm \p_{\bar l}\p_m U
 + D_+ X^\plb D_- \T^- \p_{\bar l}\p_- U}
$$ +  D_+\T^+ D_- X^{-l} \p_+ \p_l U
 +  D_+\T^+ D_- \T^- \p_+ \p_- U$$
where $T_0$ is the stress tensor in a 
flat background of \twostress, 
$\p_{\bar l} = {\p\over{\p X^\plb}}$, 
$\p_l = {\p\over{\p X^\ml}}$ and  
$\p_\pm = {\p\over{\p \T^\pm}}$.

Notice that since $W^\pm$ and $\bar W^\pm$ do not appear in the 
interaction term in 
\actfive , their equations of motion imply the same equations on 
$\T^\pm ~ \bar \T^\pm$ that we had in flat space \eqmotion .
 These implied, in 
particular, that we could use an $N=2$ superconformal transformation
to set them to 
$\T^\pm=\k^\pm$
and $\Tb^\pm=\bar\k^\pm$. 
So to find which choice of $U$ corresponds to which RR 
background one can go to lightcone gauge, do the superspace
integral in \actfive,
and compare with light-cone GS vertex operators.
Alternatively, 
one can use the field redefinition to RNS variables of \twistrns\ and
compare with the covariant RNS vertex operators of Friedan, Martinec
and Shenker \fms. Since Ramond-Ramond 
vertex operators contain an odd number
of unbarred and barred fermions, one finds that 
\eqn\rrback{ \eqalign{
U= &U_{++} (X^\ml,X^\plb) \T^+\Tb^+ + U_{+-}(X^\plb,X^\ml)
\T^+\Tb^- +
\cr ~&~~~~~~~ U_{-+}(X^\plb,X^\ml)\T^-\Tb^+ +
U_{--}(X^\ml,X^\plb)\T^-\Tb^-  ~.}}
describes the RR backgrounds that we are going to be interested in.\foot{
There are other light-cone RR backgrounds that
are described by the vertex operators
$\int d^2 z\int d^4 \kappa  
D_- X^{-[l} {\bar D}_-X^{- j] } 
\T^+ \T^- \Tb^+ \Tb^- (D_-\T^- \bar D_-\Tb^-)^{-1} $.
Since these other RR backgrounds will not preserve
any target-space supersymmetry of the type we are considering, 
we do not consider them any further.  }

Besides the maximally supersymmetric Ramond-Ramond plane wave background 
of \blau ,
there are other special Ramond-Ramond plane wave 
backgrounds which preserve less supersymmetries. 
As shown in \mtwo, these
backgrounds
are described by either a real harmonic function $V(X^\plb,X^\ml)$ which
preserves at least two supersymmetries
or a holomorphic function $Y(X^\ml)$ which preserves at least
four supersymmetries. For example, the maximally supersymmetric 
plane wave is described by $Y(X^\ml) = \d_{lm} X^{-l} X^{-m}.$
In the case of a flat transverse background, it will now be shown how
to describe these plane wave backgrounds as exact N=2 superconformal field
theories using the U(4) formalism. Note that 
``exact'' superconformal invariance will always
mean vanishing of the $\beta$-function
to all perturbative orders in $\alpha'$, and possible non-perturbative
contributions will not be discussed here. In \refs{\amati,\horsteif}
 an argument
was presented for the all order conformal invariance of plane 
waves with {\it constant} field strengths. Our argument will 
also cover certain non-constant field strengths.

By replacing $\int d^2 z d^4\k ~X^\plb X^\ml$ 
with 
$\int d^2 z d^4 \k ~K(X^\plb,X^\ml)$ in $S_0$ where
$K(X^\plb,X^\ml)$ is a Ricci-flat Kahler potential, the 
plane wave backgrounds in a flat transverse background are
easily generalized to a curved transverse background. In this case,
however, the acion is not an exact N=2 superconformal
field theory because of the usual four-loop divergences in the
N=(2,2) non-linear sigma model \ref\grisaru{M. Grisaru, A. van
de Ven and D. Zanon, {\it Two-Dimensional Supersymmetric Sigma Models
on Ricci Flat Kahler Manifolds are Not Finite}, Nucl. Phys. B277
(1986) 388.}.

\subsec{Plane wave background with real harmonic function}

Besides the sixteen light-cone supersymmetries, 
there are four spacetime supersymmetry transformations that are
simply realized in this formalism. As discussed in \susytr,
these are  generated by spinors
that are singlets under $SU(4) \in SO(8)$ and act by shifts as
\eqn\susyone{\T^\pm \to \T^\pm +\e^\pm,\quad \Tb^\pm \to \Tb^\pm +\bar\e^\pm}
where $\e^\pm$ and $\bar\e^\pm$ are constant parameters. 
Although the transformations of \susyone\ naively anticommute, one
can see from the definition of $X^+$ in \restr\ that their anticommutator
generates a constant shift in $X^+$.
The supersymmetries of \susyone\ are
generically broken in the plane wave background
of \actfive, however,
there are special choices of $U$ which preserve either two or four
of these symmetries. For example, 
\eqn\realf{U=  (\T^- - \T^+)(\Tb^- -\Tb^+)
V(X^\plb,X^\ml) }
is a Ramond-Ramond plane wave background which is invariant under
the two supersymmetries in \susyone\ generated by 
 $\e^+ =\e^- $ and $\bar\e^+ = \bar\e^-$.
It will now be argued that this supersymmetric
background is an exact solution of
superstring theory when $V$ is harmonic, i.e. that
\eqn\actsix{S= S_0 +\int d^2 z d^4 \k ~ 
(\T^- - \T^+)(\Tb^- - \Tb^+) V(X^\plb,X^\ml)}
is an exact N=2 superconformal field theory if 
$\p_{\bar l}\p_l V =0$.

To prove this, note that since $S_0$ is free and since the
interaction vertex does not involve $[W^\pm,\bar W^\pm]$, the
fields $[\T^\pm,\Tb^\pm]$ can be set equal to their background
values in the interaction term. It is easy to check that 
Feynman diagrams involving a single interaction vertex are
free of divergences if $\p_{\bar l}\p_l V=0$. And Feynman
diagrams involving more than one interaction vertex are free of
divergences since, by power counting, the only possible divergences
could come from terms involving no derivatives on the background
variables. But since $[(\T^+ -\T^-)(\Tb^+ - \Tb^-)]^2 =0$,
there are no such terms.

\subsec{Plane wave background with holomorphic function}

If $Y(X^\ml)$ is a holomorphic function,
the action
\eqn\actseven{S= S_0 +\int d^2 z d^4 \k ~ 
(Y(X^\ml) \T^+\Tb^+  +\bar Y(X^\plb)\T^-\Tb^-)}
is no longer invariant under
the transformations of \susyone. However, if one defines
$W^\pm$ and $\bar W^\pm$ to transform as 
\eqn\susytwo{
\d W^+= \bar Y(X^\plb)\bar\e^-,\quad
\d W^-=  Y(X^\ml)\bar\e^+,\quad
\d \bar W^+= - \bar Y(X^\plb)\e^-,\quad
\d \bar W^-= -  Y(X^\ml)\e^+,}
the invariance under all four transformations is restored. Note
that since $Y(X^\ml)$ is holomorphic, the transformation of \susytwo\
preserves the constraints of \chiralW.

To show that \actseven\ is an exact N=2 superconformal field theory,
first set $[\T^\pm,\bar\T^\pm]$ to their background values in
the interaction vertices. Feynman diagrams involving a single
interaction vertex are zero since $Y(X^\ml)$ is holomorphic. 
For more than one interaction vertex, the only possible divergences come
from contractions between $X^\ml(z_1,\k_1)$ and $X^\plb(z_2,\k_2)$
in the vertices $Y(X^\ml)\T^+\Tb^+(z_1,\k_1)$ and
$\bar Y(X^\plb)\T^-\Tb^-(z_2,\k_2)$. Using standard 
superspace rules in momentum space \ref\supers{S.J. Gates, M. Grisaru,
M. Rocek and W. Siegel, {\it Superspace or One Thousand and One Lessons
in Supersymmetry}, Front. Phys. 58 (1983) 1, hep-th/0108200.}, each such
contraction is proportional to
\eqn\propag{
D_{1 +}\bar D_{1 +} D_{2 -}\bar D_{2 -} (\k_1 -\k_2)^2 (\bar\k_1 -\bar\k_2)^2.}
Integrating by parts, $D_{1 +}$ can be pulled off one of the
contractions of \propag. Since all other contractions are annihilated
by $D_{1 +}$, this $D_{1 +}$ derivative can only act on the background
variables. But by power counting, divergences cannot come from terms
involving derivatives on the background variables, so the action of
\actseven\ is an exact N=2 superconformal field theory.
Note that if $Y$ had depended on $X^\plb$, this argument would not
work since contractions between 
$X^\plb(z_1,\k_1)$ and $X^\ml(z_2,\k_2)$ are not annihilated by
$D_{1 +}$.

\vskip 15pt
{\bf Acknowledgements:}
We would like to thank M. Green, C. Hull, L. Maoz, 
H. Ooguri, 
J. Russo, A. Sen and A. Tseytlin
for useful conversations and
the Newton Institute for their hospitality where this work
was done.  NB would like to thank
CNPq grant 300256/94-9, Pronex grant 66.2002/1998-9, 
and FAPESP grant 99/12763-0 for partial financial support.
This research was partially conducted during the period that NB
was employed by the Clay Mathematics Institute as a CMI Prize Fellow.
JM was supported by grant DE-FG02-90ER40542.

\vskip 15pt

\newsec{Appendix: Green-Schwarz Light-Cone Interaction Point Operators}

To compute scattering amplitudes in light-cone gauge, one
needs to introduce light-cone GS operators at the 
interaction points of the light-cone string diagram\ref\intp
{M.B. Green and J.H. Schwarz, {\it Superstring Interactions},
Nucl. Phys. B218 (1983) 43\semi S. Mandelstam, {\it Interacting 
String Picture of the Fermionic String}, Prog. Theor. Phys. Suppl.
86 (1986) 163.}.
The simplest way to write this operator is 
\eqn\lco{|F(z)|^2 = |\p x^\plb \psi^\ml +\p x^\ml \psi^\plb|^2}
where $(\psi^\ml,\psi^\plb)$ is a fermionic 
$SO(8)$ vector
which is constructed as a spin field from the GS $SO(8)$
spinor $(s^\pl, s^\mlb)$.
As in the light-cone RNS formalism
\ref\threes{S. Mandelstam,
{\it Interacting String Picture of the Neveu-Schwarz-Ramond
Model}, Nucl. Phys. B69
(1974) 77.}, this interaction point
operator is necessary for preserving $SO(9,1)$ Lorentz
invariance and it is easy to see that the GS and RNS
light-cone operators are related by SO(8) triality
that maps RNS spin fields into GS spinors and maps
RNS vectors into GS spin fields.

As is well-known, the RNS light-cone operator 
$F= \p x \cdot \psi$ comes from integration over 
the fermionic N=1 supermoduli for the worldsheet gravitino
which couples to the fermionic stress tensor 
$\p x\cdot \psi$ \fms.
If one describes the light-cone string diagram as
an N=1 superconformal map using the coordinates
$[\rho (z,\k), \xi(z,\k)]$ where $[z,\k]$ parameterizes
the complex N=1 superplane, N=1 superconformal implies
that $D_\xi=
{\p\over{\p\xi}} +\xi \p_\rho$ is proportional to
$D_\k=
{\p\over{\p\k}} +\k \p_z$, which implies that 
$\xi = D_\k\rho (\p_z \rho)^{-\half}$.
In this supersheet description, the moduli of the worldsheet gravitino
are described by the value of $\xi(z_b)$ at the interaction points $z_b$ where
$\p_z\rho|_{z_b}=0$. For example, for $N$-point tree amplitudes with the
external vertex operators located at $(z_r,\k_r)$ in the complex superplane, 
$$\rho(z,\k)=
\sum_{r=1}^N P_r^+ \log(z-z_r-\k\k_r),\quad
\xi(z,\k)=\sum_r {{P_r^+(\k-\k_r)}\over{z-z_r}}(
\sum_r {P_r^+\over{z-z_r-\k\k_r}})^{-\half},$$
and the $N-2$ gravitino moduli $\xi(z_b)$ are proportional to
$\sum_r {P_r^+\k_r\over{z_b -z_r}}$ where $z_b$ 
are the zeros of $\sum_r {P_r^+\over{z -z_r}}.$ 
So integrating over $\xi(z_b)$ has the same effect as introducing
light-cone interaction point operators and allows light-cone RNS
amplitudes to be expressed as correlation functions on N=1 super-Riemann
surfaces \ref\nb{N. Berkovits,
{\it Calculation of Scattering Amplitudes for the Neveu-Schwarz
Model using Supersheet Functional Integration}, 
Nucl. Phys. B276 (1986) 650\semi N. Berkovits, {\it Supersheet
Functional Integration and the Interacting Neveu-Schwarz String},
Nucl. Phys. B304 (1988) 537.}.

For example, for tree amplitudes, 
\eqn\treerns{A= \prod_{b=2}^{N-2} \int d^2 (\rho(z_b)-\rho(z_1))
\langle \prod_{r=1}^N V_r (z_r) \prod_{b=1}^{N-2} |F(z_b)|^2\rangle }
\eqn\trtwo{= 
\prod_{b=2}^{N-2} \int d^2 (\rho(z_b)-\rho(z_1)) \prod_{b=1}^{N-2}
\int d^2 \xi(z_b) 
\langle \prod_{r=1}^N V_r (z_r,\k_r) \rangle }
\eqn\trthree{= 
\prod_{r=2}^{N-2} \int d^2 z_r \prod_{r=1}^{N-2} 
\int d^2 \k_r |M(z_r,\k_r, P_r^+)|^2 \langle \prod_{r=1}^N
V_r(z_r,\k_r)\rangle,}
where $V_r$ are the light-cone vertex operators
and
$M(z_r,\k_r,P_r^+)$ is an overall measure factor that comes from the
Jacobian ${{\prod_b \p\rho(z_b) \p\xi(z_b)}\over{\prod_r \p z_r\p \k_r}}$
and from the anomalous transformation of the partition function under
the superconformal transformation $\rho(z,\kappa)$.
Remarkably, one can show that
the measure factor simplifies to $M=(z_N-z_1)(z_{N-1}-z_1)$,
which is the usual factor that we get in the covariant $N=1$ 
formulation \fms .
Since \trthree\ contains no singularities when 
interaction points collide, writing the RNS light-cone amplitude in terms
of N=1 superconformal correlation functions resolves the problems of
light-cone contact terms and simplifies comparison with computations 
using the standard Lorentz covariant RNS approach.
In other words, the transformation from light-cone coordinates 
$[\rho(z_b),\xi(z_b)]$ to superplane coordinates $[z_r,\k_r]$ between
\trtwo\ and \trthree\ is
valid up to the usual surface terms associated with integration over
supermoduli \ref\ambig{J. Atick, J. Rabin and A. Sen, {\it
An Ambiguity in Fermionic String
Perturbation Theory}, Nucl. Phys. B299 (1988) 279.}. 
Including this surface term provides an analytic continuation
of the scattering amplitude which automatically includes all light-cone
contact terms.

As was shown ten years ago \het, a similar method can be used for treating
the GS light-cone operator of \lco.
However, instead of integrating over N=1 worldsheet supermoduli, one
needs to integrate over N=2 worldsheet supermoduli. To see this, first
write the GS interaction point operator of \lco\  as 
\eqn\lcotwo{F(z) = \lim_{z\to z_b} (\p x^\plb s^\pl(z) \S^-(z_b) +
\p x^\ml s^\mlb(z) \S^+(z_b)) (z-z_b)^\half}
where $\S^+$ and $\S^-$ are two components of an $SO(8)$ antichiral spinor
constructed as a spin field from the $SO(8)$ chiral spinor $(s^\pl,s^\mlb)$.
Note that under $SU(4)\times U(1)$, 
the eight components of an antichiral $SO(8)$
spinor transform as $(6,0)$, $(1,-1)$ and $(1,+1)$ representations, and
$\S^-$ and $\S^+$ are defined as the $(1,-1)$ and $(1,+1)$ components. 
Using the spin field OPE's that $s^\mlb(z)\S^+(0)\to z^{-\half}\psi^\plb$
and 
$s^\pl(z)\S^-(0)\to z^{-\half}\psi^\ml$, one easily sees that \lcotwo\
is equivalent to \lco.

Since $\p x^\plb s^\pl$ and $\p x^\ml s^\mlb$ are the N=2 fermionic
stress tensors implied by the action of \acttwo\
and $\S^\pm$ are the N=2 spectral flow operators,
one can understand the GS light-cone operator of \lcotwo\ as coming
from integration over N=2 supermoduli combined with appropriately chosen
$U(1)$ twists. Although the specific combination of N=2 fermionic stress
tensors and spectral flow operators appearing in \lcotwo\ might seem
strange, it is explained by describing the light-cone string diagram
as an N=2 superconformal map from the complex N=2 superplane to the 
string diagram. Using $[\rho(z,\k^+,\k^-), \xi^+
(z,\k^+,\k^-),\xi^-(z,\k^+,\k^-)]$ as this superconformal map where
$[z,\k^+,\k^-]$ parameterize the N=2 superplane, N=2 superconformal implies
that $D_{\xi^\pm}=
{\p\over{\p\xi^\pm}} +\xi^\mp \p_\rho$ is proportional to
$D_{\k^\pm}=
{\p\over{\p\k^\pm}} +\k^\mp \p_z$, which implies that 
\eqn\ntwo{\xi^+ = (D_{\k^-}\rho) ~(D_{\k^-} D_{\k^+}\rho)^{-\half} 
f(z+\k^-\k^+),\quad
\xi^- = (D_{\k^+}\rho) ~(D_{\k^+} D_{\k^-}\rho)^{-\half} 
f^{-1}(z-\k^-\k^+)}
where $f(z)$ is an arbitrary function associated with $U(1)$
twists.

Note that $\xi^\pm$ must be a periodic function of $z$ in order that
the GS fermions $(s^\mlb, s^\pl)$ are periodic in the string diagram.
This means that the function $f(z)$ in \ntwo\ must be chosen such that
it contains square-root cuts at the same locations as the square-root cuts 
in $(\p_z\rho)^\half$. So if $\p_z\rho$ has zeros at $z=z_b$ and poles
at $z=z_r$, 
\eqn\ffdef{f = c\sqrt{\prod_b (z-z_b)^{N_b} \prod_r (z-z_r)^{N_r}}}
where $c$ is a constant and $(N_b,N_r)$ are integers. Furthermore,
the boundary condition that $\xi^+$ and $\xi^-$ have at most poles
at $z=z_b$ implies that $N_b$ is either $\pm 1$. 
The choice of $N_r$ is fixed by the boundary conditions on
the $r^{th}$ external string, e.g. $N_r=1$ implies that the 
$s^\mlb$ zero modes annihilate
the ``ground state'' whereas $N_r=-1$ implies that the
$s^\pl$ zero modes annihilate
the ``ground state''. However, the choice of $N_b$ is unfixed by
external boundary conditions, which means that all $2^B$ possible choices
of $N_b=\pm 1$ are allowed where $B$ is the number of interaction points.
Each such choice corresponds to an individual term in the light-cone
operator of \lcotwo. For example, if $N_b=+1$ for $b=1$ to $H$ and
$N_b=-1$ for $b=H+1$ to $B$, then $\xi^+$ has poles at $z_b$ for $b=1$ to $H$
and $\xi^-$ has poles at $z_b$ for $b=H+1$ to $B$.
So the term $\prod_{b=1}^H \p x^\ml s^\mlb \S^+ \prod_{b=H+1}^B 
\p x^\plb s^\pl \S^-$ in \lcotwo\ is obtained by integrating over
$\prod_{b=1}^H\int d\xi^+(z_b)\prod_{b=H+1}^B\int d\xi^-(z_b)$
where $\xi^\pm(z_b)$ signifies the residue of the pole at $z=z_b$.

As in the light-cone RNS supersheet formalism, this superconformal method
allows light-cone GS amplitudes to be expressed as correlation functions
on super-Riemann surfaces. For example, for the $N$-point tree amplitude
described by the map $\rho= \sum_{r=1}^N P_r^+ \log(z-z_r-\k^+\k^-_
r-\k^-\k_r^+)$, 
\eqn\treegs{A= \prod_{b=2}^{N-2} \int d^2 (\rho(z_b)-\rho(z_1))
\langle \prod_{r=1}^N V_r (z_r) \prod_{b=1}^{N-2} |F(z_b)|^2\rangle }
\eqn\trgstwo{= 
\prod_{b=2}^{N-2} \int d^2 (\rho(z_b)-\rho(z_1)) ~|\sum_{K=1}^{2^{N-2}}
\prod\int d\xi^+(z_b)\prod\int d\xi^-(z_b)|^2~
\langle \prod_{r=1}^N V_r (z_r,\k^+_r,\k^-_r) \rangle }
$$= 
\prod_{r=2}^{N-2} \int d^2 z_r \prod_{b=1}^{N-2}
\int d^2 \xi^+(z_b)\int d^2 \xi^-(z_b)~
|\sum_{K=1}^{2^{N-2}}\prod_{b=1}^H \xi^-(z_b)
\prod_{b=H+1}^{N-2}\xi^+(z_b)|^2  ~
\langle \prod_{r=1}^N
V_r(z_r,\k_r)\rangle$$
\eqn\trgsthree{= 
\prod_{r=2}^{N-2} \int d^2 z_r \prod_{r=1}^{N-2}
\int d^2 \k_r^+~ d^2 \k_r^- ~|M(z_r,\k_r^\pm, P_r^+)|^2 ~
\langle \prod_{r=1}^N
V_r(z_r,\k_r)\rangle}
where 
$\sum_{K=1}^{2^{N-2}}$ sums over the $2^{N-2}$ different possible
boundary conditions at the interaction points and
$M(z_r,\k^\pm_r,P_r^+)$ is an overall measure factor that comes from the
Jacobian 
$$
{{\prod_b \p\rho(z_b) \p\xi^+(z_b)
\p\xi^-(z_b)}
\over{\prod_r
 \p z_r \p \k^+_r\p\k^-_r}} 
\sum_{K=1}^{2^{N-2}}(\prod_{b=1}^H \xi^-(z_b)
\prod_{b=H+1}^{N-2}\xi^+(z_b))$$
and from the anomalous transformation of the partition function under
the superconformal transformation $\rho(z,\k^+,\k^-)$.
Unfortunately, unlike the RNS measure factor, the GS measure factor $M$
has a complicated form which has prevented \trgsthree\ from being used
to obtain super-Koba-Nielsen-like formulas for GS tree amplitudes.
Nevertheless, it can be argued that $M$ has no singularities when
interaction points collide. So as in the RNS amplitude of \trthree,
expressing the GS amplitude in terms of the N=2 superplane coordinates
$[z_r,\k_r^\pm]$
resolves the problem of light-cone contact terms by including a surface
term which provides an appropriate analytic continuation of the 
scattering amplitude.

\listrefs

\end